\documentclass[aps,prc,preprint,amsfonts,amsmath,superscriptaddress,floatfix,nofootinbib]{revtex4-1}
\usepackage{graphicx,epsfig,color,amsmath, booktabs}
\DeclareMathAlphabet{\mathpzc}{OT1}{pzc}{m}{it}

\newcommand{\beqy}{\begin{eqnarray}}
\newcommand{\eeqy}{\end{eqnarray}}

\newcommand{\U}{\mathcal{U}^{(q)}}
\newcommand{\V}{\mathcal{V}^{(q)}}

\newcommand{\Uconj}{\mathcal{U}^{(q)*}}
\newcommand{\Vconj}{\mathcal{V}^{(q)*}}
\newcommand{\Veff}{\mathbb{V}_q}
\newcommand{\E}{\mathfrak{E}^{(q)}}
\newcommand{\Etilde}{\breve{\mathfrak{E}}^{(q)}}
\newcommand{\h}{h_q}
\newcommand{\D}{\Delta_q}
\newcommand{\Dtilde}{\breve{\Delta}_q}
\newcommand{\Dk}{\Delta^{(q)}_k}
\newcommand{\Dkconj}{\Delta^{(q)*}_k}

\newcommand{\f}{f^{(q)}}
\newcommand{\fPlus}{f^{(q)}_{\pmb{k}+\pmb{Q_q}}}
\newcommand{\fMoins}{f^{(q)}_{-\pmb{k}+\pmb{Q_q}}}

\newcommand{\eps}{\varepsilon^{(q)}}
\newcommand{\epstilde}{\breve{\varepsilon}^{(q)}}
\newcommand{\sqeps}{\varepsilon^{(q)2}}
\newcommand{\sqepstilde}{\breve{\varepsilon}^{(q)2}}

\begin{document}

\title{Entrainment effects in neutron-proton mixtures within the nuclear energy-density functional theory. II. Finite temperatures and arbitrary currents.}

  \author{V. Allard}
 \affiliation{Institute of Astronomy and Astrophysics, Universit\'e Libre de Bruxelles, CP 226, Boulevard du Triomphe, B-1050 Brussels, Belgium}

 \author{N. Chamel}
 \affiliation{Institute of Astronomy and Astrophysics, Universit\'e Libre de Bruxelles, CP 226, Boulevard du Triomphe, B-1050 Brussels, Belgium}

\date{\today}

\begin{abstract}
Mutual entrainment effects in hot neutron-proton superfluid mixtures are studied in the framework of the self-consistent nuclear energy-density functional theory. The local mass 
currents in homogeneous or inhomogeneous nuclear systems, which we derive from the time-dependent Hartree-Fock-Bogoliubov equations at finite temperatures, are shown to have the same formal expression as the ones we found earlier in the absence of pairing at zero temperature. 
Analytical expressions for the entrainment matrix are obtained for application to superfluid neutron-star cores. Results are compared to those obtained earlier using Landau's theory.  
Our formulas, valid for arbitrary temperatures and currents, are applicable to various types of functionals including the Brussels-Montreal series for which unified equations of state have been already calculated, thus laying the ground for a fully consistent microscopic description of superfluid neutron stars. 
\end{abstract}

\maketitle

\section{Introduction}

Formed from the gravitational core-collapse of massive stars during supernova explosions, neutron stars are initially very hot but rapidly cool down by emitting neutrinos. Their very dense matter is thus expected to undergo various phase transitions~\cite{blaschke2018}. In particular, the core of a mature neutron star is thought to contain a neutron-proton superfluid mixture (see, e.g., Ref.~\cite{chamel2017} for a recent review). Because a superfluid can flow without resistance and carries no heat, the dynamics of a neutron star must be described by three distinct components at least: the neutron superfluid, the proton superconductor, and the normal fluid. Due to strong nuclear interactions, neutrons and protons cannot flow completely independently and are mutually entrained, similarly to superfluid $^{4}$He-$^{3}$He mixtures~\cite{andreev1976}.

Although a fully relativistic treatment is required for an accurate description of the global dynamics of neutron stars, the flows of neutrons and protons remain essentially nonrelativistic at the nuclear length scales of interest here (at such scales, spacetime curvature can also be safely ignored, as shown e.g. in Ref.~\cite{glendenning2000}). Therefore, we shall consider here nonrelativistic superfluid dynamics. 
The mass current $\pmb{\rho_q}$ of one nucleon species ($q=n,p$ for neutron, proton) is expressible as a combination of the ``superfluid velocities'' (momenta per unit mass) $\pmb{V_q}$ of both species and of the normal velocity $\pmb{v_{N}}$ of thermal excitations, as~\cite{gusakov2005} 
\beqy\label{eq:entrainment1}
\pmb{\rho_n} &=& (\rho_n-\rho_{nn}-\rho_{np})\pmb{v_{N}} + \rho_{nn} \pmb{V_{n}}+\rho_{np} \pmb{V_{p}}\, , \\  
\label{eq:entrainment2}
\pmb{\rho_p} &=& (\rho_p-\rho_{pp}-\rho_{pn})\pmb{v_{N}} + \rho_{pn} \pmb{V_{n}}+\rho_{pp} \pmb{V_{p}}\, .
\eeqy
It follows from Eqs.~\eqref{eq:entrainment1} and \eqref{eq:entrainment2} that the normal fluid carries a momentum density 
given by 
\beqy
\pmb{\rho_n} +\pmb{\rho_p} - \rho_{n} \pmb{V_{n}}-\rho_{p} \pmb{V_{p}}  =(\rho_n-\rho_{nn}-\rho_{np})(\pmb{v_{N}}- \pmb{V_{n}})+(\rho_p-\rho_{pp}-\rho_{pn})(\pmb{v_{N}}-\pmb{V_{p}})\, .
\eeqy
As shown in Ref.~\cite{gusakov2006}, the relativistic entrainment matrix, denoted by $Y_{qq'}$ and relating the nucleon four-currents to the superfluid four-velocities, can be inferred from its nonrelativistic counterpart $\rho_{qq'}$.

The (symmetric) entrainment matrix $\rho_{qq^\prime}$ is a key microscopic ingredient for modeling the dynamics of neutron stars, see, e.g. Refs.~\cite{anderssoncomer2007,chamel2017,graber2017,haskell-sedrakian2018} and references therein. It should be stressed that the entrainment matrix itself may depend on the superfluid flows, and this could play an important role on the dynamics of neutron stars~\cite{gusakov2013}. 
Although observed neutron stars are usually cold, meaning that their internal temperature $T$ is much lower than the Fermi temperatures $T_{\text{F}q}$, thermal effects may still be significant for the superfluid dynamics since the associated critical temperatures $T_{\text{c}q}$ are typically much lower than $T_{\text{F}q}$. In particular, it has been recently shown that the temperature-dependence of the entrainment matrix may have implications for neutron-star oscillations~\cite{dommes2019,kantor2020}. 

The entrainment matrix of a neutron-proton superfluid mixture at finite temperatures
was first calculated in Ref.~\cite{gusakov2005} within Landau's theory of Fermi liquids suitably extended 
to deal with superfluid systems~\cite{larkin1963,leggett1965}. Calculations were performed assuming $V_q$
are small compared to the corresponding Fermi velocities $V_{\text{F}q}$, and thus considering first-order current perturbations of the static state. 
The same approach was later extended to relativistic mixtures allowing  
for the presence of hyperons~\cite{gusakov2009b}. Landau parameters were calculated using a relativistic $\sigma-\omega-\rho$ mean-field model including scalar self-interactions but ignoring pairing. Numerical results for the relativistic entrainment matrix were obtained using the Lagrangian parametrization of Ref.~\cite{glendenning1985}, and employing the same empirical fits for the dependence on the critical temperatures as in Ref.~\cite{gusakov2005}. 
More recently, some nonlinear effects of the superfluid flows have been taken into account in the calculations of the 
neutron-proton entrainment matrix~\cite{leinson2018}. 
However, Landau's theory, on which all these studies rely, is not self-consistent, as 
emphasized in Ref.~\cite{gusakov2005}.  
Moreover this approach cannot be easily transposed to inhomogeneous systems such as the inner
crust of neutron stars, where superfluid neutrons coexist with nuclear clusters.

We have recently calculated the entrainment matrix of neutron-proton mixtures at low temperatures
~\cite{ChamelAllard2019} within the self-consistent nuclear energy-density functional
theory~\cite{duguet2014}. 
This theory has been already applied by different groups to determine the equation of state of cold dense matter using the same
functional in all regions of neutron stars (crust, mantle, and core), thus ensuring a unified and
thermodynamically consistent treatment, see, e.g. Refs.~\cite{douchinhaensel01,potekhin2013,sharma2015,pearson2018,mutafchieva2019,pearson2020,mondal2020}. Moreover, this theory has been also employed to compute the properties of superfluid neutrons in neutron-star crusts (see, e.g., Refs.~\cite{monrozeau2007, chamel2010,pastore2015} for pairing gaps, critical temperatures, and specific heat;  Refs.~\cite{chamel2017b,watanabe2017,kashiwaba2019} for superfluid fractions) and the dynamics of quantized vortices~\cite{avogadro2007,wlazlowski2016,bulgac2019}. 
In this second paper, we extend our previous analysis to finite temperatures and arbitrary currents.  The dependence of the entrainment matrix on temperature and superflows are taken into account fully self-consistently within the time-dependent Hartree-Fock-Bogoliubov (TDHFB) method.  

After introducing the TDHFB method in Sect.~\ref{sec:TDHFB} and deriving the general expressions for the local currents and superfluid velocities valid for any (homogeneous or inhomogeneous) nuclear system, 
calculations of the  entrainment matrix in the outer core of neutron stars are presented in Sect.~\ref{sec:entrainment-TDHFB}. Landau's  approximations are also discussed. Throughout the paper, we shall ignore the small difference between the neutron and proton masses, and the nucleon mass will be denoted by $m$.

\section{Time-dependent Hartree-Fock-Bogoliubov equations}
\label{sec:TDHFB}

\subsection{Matrix formulation} 

The TDHFB method is discussed, e.g. in the classical textbook of  Ref.~\cite{blaizot1986}. 
The energy $E$ of a nucleon-matter element of volume $V$ is expressed as a function of the one-body density matrix $n_q^{ij}$ and pairing tensor $\kappa_q^{ij}$ defined by the following thermal averages of the creation and destruction operators, $c_q^{i\dagger}$ and $c_q^i$ (using the symbol $\dagger$ for Hermitian conjugation), for nucleons of charge type $q$ in a quantum state $i$ (using the symbol $*$ for complex conjugation): 
\begin{equation}\label{eq:density-matrix-def}
    n_q^{ij}=\langle c_q^{j\dagger} c_q^i \rangle = n^{ji*}_q \, ,
\end{equation}
\begin{equation}
    \kappa_q^{ij}=\langle c_q^j c_q^i \rangle=-\kappa_q^{ji} \, .
\end{equation}
Introducing the generalized Bogoliubov transformation\footnote{In Ref.~\cite{blaizot1986}, the matrices were denoted by $X_j^i=\mathcal{U}_{ij}$ and $Y_j^i=\mathcal{V}_{ij}$.}
\beqy
\begin{pmatrix} b_q^i \\ b_q^{i\dagger}\end{pmatrix} = \sum_{j}\begin{pmatrix}\Uconj_{ij} & \Vconj_{ij} \\ \V_{ij} & \U_{ij}\end{pmatrix}\begin{pmatrix} c_q^j \\ c_q^{j\dagger}\end{pmatrix} \, , 
\eeqy
such that $\langle b_q^{j\dagger} b_q^i \rangle = \delta^{ij} f^{(q)}_i$ ($\delta^{ij}$ is the Kronecker's symbol) and $\langle b_q^j b_q^i \rangle=\langle b_q^{j\dagger} b_q^{i\dagger} \rangle = 0$, where $b_q^{i\dagger}$ and $b_q^i$ are creation and destruction operators of a quasiparticle in a quantum state $i$,  
the TDHFB equations, which formally take the same form at any temperature, 
can be written as~\cite{blaizot1986} 
\beqy
i\hbar\frac{\partial \U_{ki}}{\partial t}=\sum_{j}\bigl[(\h^{ij}-\lambda_q\delta^{ij})\U_{kj}+\D^{ij}\V_{kj}\bigr]\, ,
\label{eq:TDHFB-Uij}
\eeqy
\beqy
i\hbar\frac{\partial \V_{ki}}{\partial t}=\sum_{j}\bigl[-\D^{ij*}\U_{kj}-(\h^{ij*}-\lambda_q \delta^{ij})\V_{kj}\bigr] \, ,
\label{eq:TDHFB-Vij}
\eeqy
where $\lambda_q$ denotes the chemical potentials. The matrices  $h_q^{ij}$ and  $\Delta_q^{ij}$ of the single-particle Hamiltonian and the pair potential, respectively, are defined as 
\beqy\label{eq:Hamiltonian-matrix}
\h^{ij}=\frac{\partial E}{\partial n_q^{ji}}=\h^{ji*}
\, , 
\eeqy
\beqy\label{eq:pairing-matrix}
\D^{ij}=\frac{\partial  E}{\partial \kappa_q^{ij*}}=-\D^{ji}
\, .
\eeqy
The fermionic algebra of the particle operators ($c^i_q$ and $c_q^{i\dagger}$) and the quasiparticle operators ($b^i_q$ and $b_q^{i\dagger}$) yields the following identities
\beqy \label{eq:constraint1}
 \sum_k  \left(\U_{ik}\Uconj_{jk} + \V_{ik}\Vconj_{jk}\right)=\delta_{ij}\, , \qquad \sum_{k}\left(\Uconj_{ki}\U_{kj}+\V_{ki}\Vconj_{kj}\right)=\delta_{ij} ,
\eeqy 
\beqy \label{eq:constraint2}
\sum_k \left(\U_{ik}\V_{jk} + \V_{ik}\U_{jk}\right)=0\, ,  \qquad \sum_{k}\left(\Uconj_{ki}\V_{kj}+\V_{ki}\Uconj_{kj}\right)=0 \, .
\eeqy 
The one-body density matrix and the pairing tensor can be expressed in terms of the quasiparticle components as
\beqy\label{eq:DensityMatrix}
n_q^{ij} = \sum_k \left[\f_k\U_{ki}\Uconj_{kj} +(1-\f_k) \Vconj_{ki}\V_{kj}\right]\, ,
\eeqy
\beqy
\kappa_q^{ij} =\sum_k \left[\f_k\U_{ki}\Vconj_{kj} +(1-\f_k)\Vconj_{ki}\U_{kj}\right] \, .
\eeqy

\subsection{Coordinate-space formulation}

The energy $E$ is generally further assumed to depend on $n_q^{ij}$ and $\kappa_q^{ij}$ only through the following local densities and currents\footnote{The energy may be a functional of  other densities and currents. We consider here only those relevant for calculating the entrainment couplings in homogeneous nuclear matter using the most popular functionals.}: 

\noindent (i) the nucleon number density at position $\pmb{r}$ and time $t$
\beqy
\label{eq:dens-def}
n_q(\pmb{r},t) = \sum_{\sigma=\pm 1}n_q(\pmb{r}, \sigma; \pmb{r}, \sigma;t)\, ,
\eeqy
(ii) the pair density (in general a complex number) at position $\pmb{r}$ and time $t$
\beqy
\label{eq:abdens-def}
\widetilde{n}_q(\pmb{r},t) = \sum_{\sigma=\pm 1}\widetilde{n}_q(\pmb{r}, \sigma; \pmb{r}, \sigma;t)\, ,
\eeqy
(ii) the kinetic energy density (in units of $\hbar^2/2m$) at position $\pmb{r}$ and time $t$
\beqy
\label{eq:kin-def}
\tau_q(\pmb{r},t) = \sum_{\sigma=\pm 1}\int\,{\rm d}^3\pmb{r^\prime}\,\delta(\pmb{r}-\pmb{r^\prime}) \pmb{\nabla}\cdot\pmb{\nabla^\prime}
n_q(\pmb{r}, \sigma; \pmb{r^\prime}, \sigma;t)\, ,
\eeqy
(iii) and the momentum density (in units of $\hbar$) at position $\pmb{r}$ and time $t$
\beqy
\label{eq:mom-def}
\pmb{j_q}(\pmb{r},t)=-\frac{ i}{2}\sum_{\sigma=\pm 1}\int\,{\rm d}^3\pmb{r^\prime}\,\delta(\pmb{r}-\pmb{r^\prime}) (\pmb{\nabla} -\pmb{\nabla^\prime})n_q(\pmb{r}, \sigma; \pmb{r^\prime}, \sigma;t)\, ,
\eeqy
where the particle and pair density matrices in coordinate space are respectively defined by~\cite{dobaczewski1984}
\begin{equation}
n_q(\pmb{r}, \sigma; \pmb{r^\prime}, \sigma^\prime; t) = <c_q(\pmb{r^\prime},\sigma^\prime; t)^\dagger c_q(\pmb{r},\sigma; t)>\, ,
\end{equation}
\begin{equation}
\widetilde{n}_q(\pmb{r}, \sigma; \pmb{r^\prime}, \sigma^\prime ;t) = -\sigma^\prime <c_q(\pmb{r^\prime},-\sigma^\prime; t) c_q(\pmb{r},\sigma ; t)>\, , 
\end{equation}
where $c_q(\pmb{r},\sigma;t)^\dagger$ and $c_q(\pmb{r},\sigma;t)$ are the creation and destruction operators for nucleons of charge type $q$ at position $\pmb{r}$ with spin $\sigma$ at time t. Introducing single-particle basis wavefunctions $\varphi^{(q)}_i(\pmb{r},\sigma)$, these matrices can be alternatively written in terms of $n^{ij}_{q}$ and $\kappa^{ij}_{q}$ as 
\begin{equation}\label{eq:DensityMatrixCoordinateSpaceDef}
n_q(\pmb{r}, \sigma; \pmb{r^\prime}, \sigma^\prime; t)=\sum_{i,j}n^{ij}_{q}\varphi^{(q)}_i (\pmb{r},\sigma)\varphi^{(q)}_j (\pmb{r^\prime},\sigma^\prime)^{*}\, ,
\end{equation}
\begin{equation}\label{eq:AbnormalDensityMatrixCoordinateSpaceDef}
\widetilde{n}_q(\pmb{r}, \sigma; \pmb{r^\prime}, \sigma^\prime; t)=-\sigma^\prime\sum_{i,j}\kappa^{ij}_{q}\varphi^{(q)}_i(\pmb{r},\sigma)\varphi^{(q)}_j(\pmb{r^\prime},-\sigma^\prime)\, . 
\end{equation}
Examples of nuclear energy-density functionals depending on the above local densities and currents are those constructed from zero-range effective nucleon-nucleon interactions of the Skyrme type~\cite{bender03,stone2007}. The present formalism is also applicable to 
more general classes of functionals, such as those proposed in Ref.~\cite{chamel2009}. 
The pair density matrix is related to the order parameter $\Psi_q(\pmb{r},t)$ of the superfluid phase at position $\pmb{r}$ and time $t$   as follows (see, e.g., Eq.(2.4.24) of Ref.~\cite{leggett2006})
\begin{equation}\label{eq:order-param}
\Psi_q(\pmb{r},t) \equiv \widetilde{n}_q(\pmb{r}, -1; \pmb{r}, -1; t)=\widetilde{n}_q(\pmb{r}, +1; \pmb{r}, +1; t)=\frac{1}{2}\widetilde{n}_q(\pmb{r}, t)\, .
\end{equation}
The matrices (\ref{eq:Hamiltonian-matrix}) and (\ref{eq:pairing-matrix}) of the single-particle Hamiltonian and pair potential are given by 
\beqy\label{eq:Hamiltonian-matrix2}
\h^{ij}(t)&=&\int {\rm d}^3\pmb{r} \, \left[ \frac{\delta E}{\delta n_q(\pmb{r},t)}\frac{\partial n_q(\pmb{r},t)}{\partial n_q^{ji}}+\frac{\delta E}{\delta \tau_q(\pmb{r},t)}\frac{\partial \tau_q(\pmb{r},t)}{\partial n_q^{ji}}+\frac{\delta E}{\delta \pmb{j_q}(\pmb{r},t)}\cdot\frac{\partial \pmb{j_q}(\pmb{r},t)}{\partial n_q^{ji}} \right]\nonumber \\ 
&=&\sum_{\sigma}\int {\rm d}^3\pmb{r} \,  \varphi^{(q)}_i(\pmb{r},\sigma)^*\h(\pmb{r},t)\varphi^{(q)}_j(\pmb{r},\sigma)\, , 
\eeqy
\beqy\label{eq:pairing-matrix2}
\D^{ij}(t)&=& \int {\rm d}^3\pmb{r} \, \frac{\delta E}{\delta \widetilde{n}_q(\pmb{r},t)^*}\frac{\partial \widetilde{n}_q(\pmb{r},t)^*}{\partial \kappa_q^{ij*}} \nonumber \\
&=&-\sum_{\sigma}\sigma \int {\rm d}^3\pmb{r} \,  \varphi^{(q)}_i(\pmb{r},\sigma)^*\D(\pmb{r},t)\varphi^{(q)}_j(\pmb{r},-\sigma)^*\, , 
\eeqy
where
\beqy
\label{eq:Hamiltonian}
\h(\pmb{r},t)&=&-\pmb{\nabla}\cdot \frac{\hbar^2}{2 m_q^\oplus(\pmb{r},t)}\pmb{\nabla} + U_q(\pmb{r},t)-
\frac{i}{2}\biggl[\pmb{I_q}(\pmb{r},t)\cdot\pmb{\nabla}+\pmb{\nabla}\cdot\pmb{I_q}(\pmb{r},t)\biggr]\, ,
\eeqy
\beqy
\label{eq:def-fields}
\frac{\hbar^2}{2 m_q^\oplus(\pmb{r},t)}=\frac{\delta E}{\delta \tau_q(\pmb{r},t)}, \qquad U_q(\pmb{r},t)=\frac{\delta E}{\delta n_q(\pmb{r},t)},\qquad 
\pmb{I_q}(\pmb{r},t)=  \frac{\delta E}{\delta \pmb{j_q}(\pmb{r},t)}\, ,
\eeqy
\beqy
\label{eq:pair-pot}
\D(\pmb{r},t) =2\frac{\delta E}{\delta \widetilde{n}_q(\pmb{r},t)^*}\, .
\eeqy
The factor of $2$ in Eq.~\eqref{eq:pair-pot} arises from the antisymmetry of the pairing tensor $\kappa_q^{ij}$ (taking the derivative with respect to $\kappa_q^{ij*}$ is equivalent to taking the derivative with respect to $-\kappa_q^{ji*}$). Because $\pmb{I_q}$ is a vector, it must obviously depend itself on $\pmb{j_q}$. This may also be the case for all the other fields. For instance, the potential $U_q$ derived from the Brussels-Montreal functionals BSk19-26~\cite{goriely2010,goriely2013} depends on $j_n^2$,  $j_p^2$ and $\pmb{j_n}\cdot\pmb{j_p}$.

Since the energy $E$ is real, it can only depend on the pair density through its square modulus $\vert\widetilde{n}_q(\pmb{r},t)\vert^2$. The pairing potential~\eqref{eq:pair-pot} can thus be written as
\begin{equation}\label{eq:pair-pot2}
    \D(\pmb{r},t) = 2\frac{\delta E}{\delta \vert\widetilde{n}_q(\pmb{r},t)\vert^2}\widetilde{n}_q(\pmb{r},t)=4\frac{\delta E}{\delta \vert\widetilde{n}_q(\pmb{r},t)\vert^2}\Psi_q(\pmb{r},t) \, .
\end{equation}
The last equality shows that $\D(\pmb{r},t)$ has the same phase as the order parameter $\Psi_q(\pmb{r},t)$. 
Using Eq.~(\ref{eq:AbnormalDensityMatrixCoordinateSpaceDef}), the matrix elements~(\ref{eq:pairing-matrix2}) of the pairing field~\eqref{eq:pair-pot2} will thus generally take the following form: 
\begin{equation}
\D^{ij}=\frac{1}{2}\sum_{k,l}v^{(q)}_{ijkl}\kappa^{kl}_{q}\, ,
\end{equation}
with
\begin{equation}\label{eq:pairing-interaction-general}
v^{(q)}_{ijkl}\equiv 4\sum_{\sigma,\sigma^\prime}\sigma\sigma^\prime\int {\rm d}^3\pmb{r} \,  \frac{\delta E}{\delta \vert\widetilde{n}_q(\pmb{r},t)\vert^2}  \varphi^{(q)}_i(\pmb{r},\sigma)^*\varphi^{(q)}_j(\pmb{r},-\sigma)^*\varphi^{(q)}_k(\pmb{r},\sigma^\prime)\varphi^{(q)}_l(\pmb{r},-\sigma^\prime)\, .
\end{equation}
Let us note that $v^{(q)}_{ijkl}=v^{(q)*}_{klij}=-v^{(q)}_{jikl}=-v^{(q)}_{ijlk}$. 
In the traditional formulation of the TDHFB equations (see, e.g., Ref.~\cite{blaizot1986}), $v^{(q)}_{ijkl}$ represents the matrix elements of the effective (in-medium) two-body pairing interaction. 

\subsection{Mass currents and superfluid velocities}
\label{sec:micro-currents}

The TDHFB Eqs.~\eqref{eq:TDHFB-Uij} and \eqref{eq:TDHFB-Vij} can be conveniently rewritten as~\cite{blaizot1986}
\beqy
\label{eq:TDHFB1}
i\hbar \frac{\partial n_q^{ij}}{\partial t}=\sum_k \left(\h^{ik}n_q^{kj}-n_q^{ik}\h^{kj}+\kappa_q^{ik}\D^{kj*} -\D^{ik}\kappa_q^{kj*}\right) \, ,
\eeqy
\beqy
\label{eq:TDHFB2}
i \hbar \frac{\partial \kappa_q^{ij}}{\partial t}=\sum_k  \left[(\h^{ik}-\lambda_q\delta^{ik})\kappa_q^{kj}+\kappa_q^{ik}(\h^{kj*}-\lambda_q \delta^{kj})-\D^{ik}n_q^{kj*} - n_q^{ik}\D^{kj}\right]+ \D^{ij}\, .
\eeqy
As shown in Appendix~\ref{app:continuity}, Eq.~\eqref{eq:TDHFB1} can be translated in coordinate space by the same continuity equations as in the absence  of pairing (a similar conclusion was previously drawn in Ref.~\cite{gusakov2010} within Landau's theory)
\beqy\label{eq:continuity-final}
 \frac{\partial \rho_q(\pmb{r},t)}{\partial t}+\pmb{\nabla}\cdot\pmb{\rho_q}(\pmb{r},t) = 0\, ,
\eeqy
where the nucleon mass current is given by~\cite{ChamelAllard2019}
\beqy\label{eq:mass-current}
\pmb{\rho_q}(\pmb{r},t) =\frac{m}{m_q^\oplus(\pmb{r},t)} \hbar \pmb{j_q}(\pmb{r},t)+\rho_q(\pmb{r},t) \frac{\pmb{I_q}(\pmb{r},t)}{\hbar} \, .
\eeqy
The effective mass $m_q^\oplus(\pmb{r},t)$ and the vector field $\pmb{I_q}(\pmb{r},t)$ are defined by Eq.~(\ref{eq:def-fields}). As shown in our previous work~\cite{ChamelAllard2019}, the nucleon mass current can be expressed solely in terms of the momentum densities as 
\beqy\label{eq:mass-current-C}
\pmb{\rho_q}(\pmb{r},t) &=&\hbar \pmb{j_q}(\pmb{r},t)\Biggl\{1+\frac{2}{\hbar^2}\Biggl[\frac{\delta E^j_{\rm nuc}}{\delta X_0(\pmb{r},t)}-\frac{\delta E^j_{\rm nuc}}{\delta X_1(\pmb{r},t)}\Biggr]\rho(\pmb{r},t)\Biggr\}\nonumber \\ 
&&-\hbar \pmb{j}(\pmb{r},t) \frac{2}{\hbar^2}\Biggl[\frac{\delta E^j_{\rm nuc}}{\delta X_0(\pmb{r},t)}-\frac{\delta E^j_{\rm nuc}}{\delta X_1(\pmb{r},t)}\Biggr]\rho_q(\pmb{r},t)\, ,
\eeqy 
where $E^j_{\rm nuc}$ represents the nuclear-energy terms contributing to the mass currents, and we have introduced the following fields 
\begin{equation}
X_0(\pmb{r},t)=n_0(\pmb{r},t)\tau_0(\pmb{r},t) -  j_0(\pmb{r},t)^2\, ,     
\end{equation}
\begin{equation}
X_1(\pmb{r},t)=n_1(\pmb{r},t)\tau_1(\pmb{r},t) -  j_1(\pmb{r},t)^2 \, .
\end{equation} 
Let us recall that the subscripts $0$ and $1$ denote isoscalar and isovector quantities, respectively, namely sums over neutrons and protons for the former (e.g. $n_0\equiv n=n_n+n_p$) and differences between neutrons and protons for the latter (e.g. $n_1=n_n-n_p$). 

The so-called ``superfluid velocity'' of the nucleon species $q$ at position $\pmb{r}$ and time $t$ is defined by (see, e.g., Eq. (2.4.21) of Ref.~\cite{leggett2006})
\beqy\label{eq:super-vel-def}
\pmb{V_q}(\pmb{r},t)=\frac{\hbar}{2m} \pmb{\nabla}\phi_q(\pmb{r},t)\, ,
\eeqy
where $\phi_q(\pmb{r},t)$ is the phase of the associated condensate defined through the order parameter~\eqref{eq:order-param}
\beqy\label{eq:phase}
\Psi_q(\pmb{r},t)=\vert \Psi_q(\pmb{r},t)\vert \exp(i \phi_q(\pmb{r},t) )\, .
\eeqy

\section{Entrainment effects in neutron-star cores}
\label{sec:entrainment-TDHFB}

\subsection{Exact solution of the TDHFB equations in homogeneous matter}
\label{sec:TDHFB-hom}

In this section, we consider an homogeneous neutron-proton mixture with stationary nucleon currents in the normal rest frame ($\pmb{v_N}=\pmb{0}$). The latter assumption ensures that the entropy densities $s_q$ are independent of time. Recalling that~\cite{blaizot1986}
\beqy 
s_q=-\frac{k_{\rm B}}{V} \sum_i\left[f^{(q)}_i\ln f^{(q)}_i +(1-f^{(q)}_i)\ln(1-f^{(q)}_i)\right]\, ,
\eeqy 
where $k_{\rm B}$ denotes Boltzmann's constant, the quasiparticle distribution functions $f^{(q)}_i$ are therefore also independent of time. 

The single-particle wave functions are given by plane waves: 
\beqy\label{eq:PlaneWavefunctions}
\varphi_j (\pmb{r},\sigma)=\frac{1}{\sqrt{V}}\exp\left(i\pmb{k}_j\cdot\pmb{r}\right)\chi_j(\sigma)\, ,
\eeqy
where $\chi_j(\sigma)=\delta_{\sigma_j\sigma}$ denotes the Pauli spinor.
As can be seen from Eqs.~\eqref{eq:DensityMatrixCoordinateSpaceDef}, and \eqref{eq:Hamiltonian-matrix2},  the density matrix~\eqref{eq:density-matrix-def} and the single-particle Hamiltonian matrix~\eqref{eq:Hamiltonian-matrix} are both diagonal in this basis.

The superfluid velocities, which are necessarily spatially uniform and independent of time, are conveniently written as 
\beqy\label{eq:SuperfluidVelocity-def}
\pmb{V_q}\equiv\frac{\hbar \pmb{Q_q}}{m} \, .
\eeqy
This corresponds to a superfluid phase $\phi_q(\pmb{r})=2 \, \pmb{Q_q}\cdot \pmb{r}$ (modulo some arbitrary constant term without any physical consequence). 
 Inserting Eq.~\eqref{eq:PlaneWavefunctions} in Eq.~\eqref{eq:order-param} using Eqs.~\eqref{eq:abdens-def} and \eqref{eq:AbnormalDensityMatrixCoordinateSpaceDef}, 
the order parameter~\eqref{eq:order-param} thus reduces to
\beqy
\Psi_q (\pmb{r},t)=|\Psi_q(t)|\exp\left(2i\pmb{Q_q}\cdot\pmb{r}\right)=\frac{1}{4V}\sum_{i,j} \kappa_q^{ij}\exp\left[i\left(\pmb{k}_i + \pmb{k}_j\right)\cdot\pmb{r}\right](\sigma_j-\sigma_i)\, .
\eeqy
One simple choice is to consider that $\kappa^{q}_{ij}$ is non-zero only if the states $i$ and $j$ have opposite spins and wave vectors $\pmb{k}_i$ and $\pmb{k}_j$ are such that $\pmb{k}_i+\pmb{k}_j=2\pmb{Q_q}$. We can always arrange states such that $\pmb{k}_i=\pmb{k}+\pmb{Q_q}$ and $\pmb{k}_j=-\pmb{k}+\pmb{Q_q}$. For convenience, we introduce the following shorthand notation 
\beqy
k\equiv (\pmb{k}+\pmb{Q_q},\sigma)\,,\qquad\qquad \bar{k}\equiv (-\pmb{k}+\pmb{Q_q},-\sigma)\,  .
\label{eq:NonVanishingIndices}
\eeqy
These quantum numbers define the conjugate states that are paired. Indeed, it follows from the definition~(\ref{eq:pairing-matrix}) that
the only nonzero matrix elements of the pair potential are of the form $\D^{k\bar{k}}=-\D^{\bar{k}k}$. In the absence of current ($\pmb{Q_q}=\pmb{0}$), the conjugate state $\bar{k}$ is the time-reversed state of $k$. The presence of a non-vanishing current ($\pmb{Q_q}\neq\pmb{0}$) breaks the time-reversal symmetry. The nonzero elements of the $\U$ and $\V$ matrices satisfying Eqs.~\eqref{eq:constraint1} and \eqref{eq:constraint2}  are of the form $\U_{kk}=\U_{\bar{k}\bar{k}}$ and $\V_{k\bar{k}}=-\V_{\bar{k}k}$. Substituting  $i\hbar \partial/\partial t$ by the quasiparticle energies $\E_k$, the TDHFB Eqs.~(\ref{eq:TDHFB-Uij}) and~(\ref{eq:TDHFB-Vij}) finally reduce to 
\beqy\label{eq:HFB-homogeneous}
\begin{pmatrix}
\xi^{(q)}_k & \Dk  \\ 
\Dkconj & -\xi^{(q)*}_{\bar{k}}
\end{pmatrix}\begin{pmatrix} \U_{kk} \\ \V_{k\bar{k}}\end{pmatrix}=\E_k\begin{pmatrix} \U_{kk} \\ \V_{k\bar{k}}\end{pmatrix}\, , 
\eeqy
where $\xi^{(q)}_k\equiv \epsilon^{(q)}_k-\lambda_q$ ($\epsilon^{(q)}_k$ denoting the eigenvalues of the single-particle Hamiltonian matrix) are explicitly given by 
\beqy\label{eq:xi}
\xi^{(q)}_{k}=\frac{\hbar^2}{2m_q^{\oplus}}\left(\pmb{k}+\pmb{Q_q}\right)^2 + U_q+\pmb{I}_q\cdot \left(\pmb{k}+\pmb{Q_q}\right)-\lambda_q \, .
\eeqy
The matrix elements $\Dk\equiv\D^{k\bar{k}}=-\Delta^{(q)}_{\bar k}$ are given by the following equation
\begin{equation}\label{eq:gap1}
\Dk = \frac{1}{2}\sum_l v^{(q)}_{k\bar{k}l\bar{l}}\,\kappa_q^{l\bar{l}}\, .
\end{equation}
Using Eqs.~(\ref{eq:DensityMatrix}), (\ref{eq:dens-def}), (\ref{eq:mom-def}),  and (\ref{eq:DensityMatrixCoordinateSpaceDef}), the nucleon number and momentum densities read
\begin{align}\label{eq:Density}
n_q=\frac{1}{V}\sum_{k}n_q^{kk} = 
\frac{1}{V}\sum_{k}\left[|\U_{kk}|^2 \f_k+|\V_{\bar{k}k}|^2\left(1-\f_{\bar{k}}\right)\right]\, , 
\end{align}
\begin{align}\label{eq:MomentumDensity}
\pmb{j_q}=\frac{1}{V}\sum_{k}\pmb{k}_k n_q^{kk}=
\frac{1}{V}\sum_{k}\left(\pmb{k}+\pmb{Q_q}\right)\left[|\U_{kk}|^2 \f_k+|\V_{\bar{k}k}|^2\left(1-\f_{\bar{k}}\right)\right]\, , 
\end{align}
respectively, where the quasiparticle distribution is given by~\cite{blaizot1986}
\begin{equation}\label{eq:QuasiparticleDistribution}
    \f_k = \left[1+\exp\left(\beta \E_k\right)\right]^{-1}=\frac{1}{2}\left[1-\tanh\left(\frac{\beta}{2} \E_k\right)\right]\, .
\end{equation}
where $\beta \equiv \left(k_\textrm{B} T\right)^{-1}$.

The solutions of Eq.~(\ref{eq:HFB-homogeneous}), readily obtained by diagonalizing the HFB matrix, are given by\footnote{Equation~\eqref{eq:HFB-homogeneous} actually admits two kinds of solutions corresponding to the eigenvalues $\E_{k\pm} = (\xi^{(q)}_k-\xi^{(q)}_{\bar k})/2\pm\sqrt{\sqeps_k + \vert\Dk\vert^2}$. Those associated with $\E_{k-}$ are such that the expressions \eqref{eq:UTerms} and \eqref{eq:VTerms} of $\vert\U_{kk}\vert^2$ and $\vert\V_{k\bar k}\vert^2$ are swapped. However, these solutions lead to the same values for the nucleon number density~(\ref{eq:Density}) and the momentum density~(\ref{eq:MomentumDensity}) using the fact that $\E_{k-}=-\E_{\bar k+}$. Therefore, they give the same mass current~\eqref{eq:mass-current-C} and, consequently, the same entrainment matrix.}
\beqy\label{eq:QuasiparticleEnergy}
\E_{k} = \frac{\xi^{(q)}_k-\xi^{(q)}_{\bar k}}{2} +\sqrt{\sqeps_k + \vert\Dk\vert^2}\,,
\eeqy
\beqy\label{eq:UTerms}
\vert\U_{kk}\vert^2=\frac{1}{2}\left(1+ \frac{\eps_k}{\sqrt{\sqeps_k + \vert\Dk\vert^2}}  \right)\,,
\eeqy
\beqy\label{eq:VTerms}
\vert\V_{k\bar{k}}\vert^2=\frac{1}{2}\left(1- \frac{\eps_k}{\sqrt{\sqeps_k + \vert\Dk\vert^2}}  \right)\,,
\eeqy
where 
\begin{equation}\label{eq:EpsilonTerm}
  \eps_k\equiv \frac{\xi^{(q)}_k+\xi^{(q)}_{\bar k}}{2} = \eps_{\bar k}\, .
\end{equation}
Using Eq.~\eqref{eq:xi}, we have
\begin{equation}\label{eq:Epsilon}
  \eps_k=\frac{\hbar^2}{2m_q^{\oplus}}\left(\pmb{k}^2 + \pmb{Q_q}^2\right) + U_q+\pmb{I}_q\cdot \pmb{Q_q}-\lambda_q \, ,
\end{equation}
\begin{equation}\label{eq:QuasiparticleEnergyLinear}
 \frac{\xi^{(q)}_k-\xi^{(q)}_{\bar k}}{2}=\hbar \pmb{k}\cdot \pmb{\Veff}\, ,
 \end{equation}
 where we have introduced the effective superfluid velocity 
 \begin{equation}\label{eq:Veff}
\pmb{\Veff}= \frac{\hbar}{m_q^{\oplus}}\pmb{Q_q} + \frac{\pmb{I_q}}{\hbar} = \frac{m}{m_q^{\oplus}}\pmb{V_q}+\frac{\pmb{I_q}}{\hbar}\, . 
\end{equation}
For standard Skyrme functionals, the effective mass $m_q^\oplus$ depends only on the 
nucleon densities, whereas the potential $U_q$ depends on the pairing gaps $\Dk$ 
(which in turn depend also on $\pmb{Q_n}$ and $\pmb{Q_p}$). For the extended Skyrme  
functionals proposed in Ref.~\cite{chamel2009}, the potential $U_q$ depends also 
explicitly on $j_n^2$, $j_p^2$ and $\pmb{j_n}\cdot\pmb{j_p}$. In general, the dependence of the energies~\eqref{eq:Epsilon} on the superfluid velocities may be therefore quite complicated. 

Using Eqs.~\eqref{eq:pairing-interaction-general} and \eqref{eq:PlaneWavefunctions}, it can be shown that the effective pairing interaction $v^{(q)}_{k\bar{k}l\bar{l}}$ is independent of the wave vectors, and depends only on the spins. The only nonzero elements are given by  (denoting the spins with arrows for clarity) 
\begin{equation}
v^{(q)}_{\downarrow\uparrow\downarrow\uparrow}=\frac{4}{V} \frac{\delta E}{\delta\vert \widetilde{n}_q\vert^2}<0\, , 
\end{equation}
and any other element obtained by permutation of the spin indices. It thus follows from Eq.~\eqref{eq:gap1}, that $\Dk$ is also independent of the wave vectors $\pmb{k}$ (but $\Dk$ still depends on the given wave vectors $\pmb{Q_q}$). Dropping the wave vector $\pmb{k}$ as a subscript,  introducing the pairing gap 
\begin{equation}
    \D \equiv\Delta^{(q)}_{\downarrow\uparrow} = \frac{1}{V}\int d^3\pmb{r}\, \vert \Delta_q(\pmb{r})\vert\geq 0\, ,
\end{equation}
and using Eqs.~\eqref{eq:UTerms} and \eqref{eq:VTerms}, Eq.~\eqref{eq:gap1} reduces to the gap equation 
\begin{equation}\label{eq:GapEquation}
    \D = \frac{2}{V}\frac{\delta E}{\delta\vert \widetilde{n}_q\vert^2} \sum_{\pmb{k}} \frac{\D}{\sqrt{\varepsilon_k^{(q)2} +\D^2}}(f_k + f_{\bar{k}}-1)\, .
\end{equation}
The summation is only over the wave vectors $\pmb{k}$, the summation over the spins has been already carried out. Note that the right-hand side of this equation explicitly depends on the wave vectors $\pmb{Q_q}$ through Eqs.~\eqref{eq:QuasiparticleDistribution}, \eqref{eq:QuasiparticleEnergy}, \eqref{eq:Epsilon} and \eqref{eq:QuasiparticleEnergyLinear}.

\subsection{Entrainment matrix from the TDHFB solution} 

Using the solution of the TDHFB equations, the momentum density~\eqref{eq:MomentumDensity} can be alternatively written as 
\begin{align}\label{eq:MomentumDensity2}
\pmb{j_q}=n_q\, \pmb{Q_q} + 
\frac{1}{V}\sum_{\pmb{k}}\pmb{k} (\f_k-\f_{\bar k})\, .
\end{align}
The first term in the right-hand side coincides with the expression (29) adopted in our previous work~\cite{ChamelAllard2019}, thus demonstrating the validity of this identification. 
The second term accounts for quasiparticle excitations (note that the summation over spin states has been already carried out). Remarking that the component of $\pmb{k}$ orthogonal to $\pmb{\Veff}$ does not contribute to the sum (since $\f_k=\f_{\bar k}$ in this case), the momentum density can be expressed as 
\begin{equation}
\pmb{j_q}= \frac{m n_{q}}{\hbar}\pmb{V_q} - \frac{m_q^{\oplus}n_{q}}{\hbar}\mathcal{Y}_q\pmb{\Veff}\, , 
\end{equation}
with the $\mathcal{Y}_q$ function defined as
\beqy\label{eq:def-calYq}
\mathcal{Y}_q (T,\pmb{\Veff})\equiv -\frac{\hbar}{m_q^{\oplus}n_{q}\Veff^2}\frac{1}{V}\sum_{\pmb{k}}\pmb{k}\cdot\pmb{\Veff} (\f_k-\f_{\bar k})\, .
\eeqy
Substituting Eq.~\eqref{eq:QuasiparticleDistribution} yields 
\beqy\label{eq:def-calYq2}
\mathcal{Y}_q (T,\pmb{\Veff})=\frac{\hbar}{m_q^{\oplus}n_q \Veff^2}\left[\frac{1}{V}\sum_{\pmb{k}} \frac{ \pmb{k}\cdot\pmb{\Veff} \sinh\left(\beta\hbar\pmb{k}\cdot\pmb{\Veff}\right)}{\cosh\left(\beta\E_{\pmb{k}}\right) + \cosh\left(\beta\hbar\pmb{k}\cdot\pmb{\Veff}\right)}\right]\, .\eeqy
In terms of the effective superfluid velocity~\eqref{eq:Veff}, the mass current~\eqref{eq:mass-current} reduces to
\beqy\label{eq:Mass-Current-Def2}
\pmb{\rho_q} = \rho_q \left(1-\mathcal{Y}_q\right)\pmb{\Veff}\, ,
\eeqy
while the momentum density $\pmb{j_q}$ becomes 
\beqy\label{eq:Momentum-Density-Def3}
\pmb{j_q} = \frac{\rho_q}{\hbar}\left(1-\mathcal{Y}_q\right)\pmb{V_q}-\frac{m_q^{\oplus}n_q}{\hbar^2}\mathcal{Y}_q \pmb{I_q}\, .
\eeqy
Using this expression of $\pmb{j_q}$ combined with the general expression of the vector field $\pmb{I_q}$~\cite{ChamelAllard2019}
\beqy\label{eq:Ivector-def}
\pmb{I_q}=\frac{\delta E^j_{\rm nuc}}{\delta\pmb{j_q}}=-2\left(\pmb{j_p}+\pmb{j_n}\right)\left(\frac{\delta E^j_{\rm nuc}}{\delta X_0}-\frac{\delta E^j_{\rm nuc}}{\delta X_1}\right)-4\pmb{j_q}\frac{\delta E^j_{\rm nuc}}{\delta X_1}
\eeqy
leads to a self-consistent system of equations for $\pmb{j_q}$ and $\pmb{V_q}$, whose solutions are 
\beqy\label{eq:Iq-vs-Vq}
\pmb{I_q}=\sum_{q'}\mathcal{I}_{qq'}\pmb{V_{q'}}\, , 
\eeqy
\begin{align}\label{eq:Inn-def}
\mathcal{I}_{nn}&=\frac{2}{\hbar}\rho_{n}  \left(1-\mathcal{Y}_{n}\right)\Theta\left[\frac{\delta E^j_{\rm nuc}}{\delta X_1}\left(\frac{8}{\hbar^2}\frac{\delta E^j_{\rm nuc}}{\delta X_0}m_p^{\oplus}n_p\mathcal{Y}_p-1\right)-\frac{\delta E^j_{\rm nuc}}{\delta X_0}\right]\, ,
\end{align}
\begin{align}\label{eq:Ipp-def}
\mathcal{I}_{pp}&=\frac{2}{\hbar}\rho_{p}  \left(1-\mathcal{Y}_{p}\right)\Theta\left[\frac{\delta E^j_{\rm nuc}}{\delta X_1}\left(\frac{8}{\hbar^2}\frac{\delta E^j_{\rm nuc}}{\delta X_0}m_n^{\oplus}n_n\mathcal{Y}_n-1\right)-\frac{\delta E^j_{\rm nuc}}{\delta X_0}\right]\, ,
\end{align}
\begin{align}\label{eq:Inp-def}
\mathcal{I}_{np}&=\frac{2}{\hbar}\rho_{p}  \left(1-\mathcal{Y}_{p}\right)\Theta\left(\frac{\delta E^j_{\rm nuc}}{\delta X_1}-\frac{\delta E^j_{\rm nuc}}{\delta X_0}\right)\, ,
\end{align}
\begin{align}\label{eq:Ipn-def}
\mathcal{I}_{pn}&=\frac{2}{\hbar}\rho_{n}  \left(1-\mathcal{Y}_{n}\right)\Theta\left(\frac{\delta E^j_{\rm nuc}}{\delta X_1}-\frac{\delta E^j_{\rm nuc}}{\delta X_0}\right)\, ,
\end{align} 
\begin{align}\label{eq:Theta-def}
\Theta &\equiv\left[1-\frac{2}{\hbar^2}\left(\frac{\delta E^j_{\rm nuc}}{\delta X_0} +\frac{\delta E^j_{\rm nuc}}{\delta X_1}\right)\left(m_n^{\oplus}n_{n}\mathcal{Y}_{n} + m_p^{\oplus}n_{p}\mathcal{Y}_{p} \right)\right.\nonumber \\
&\qquad\;\; \left. +\left(\frac{4}{\hbar^2}\right)^2\frac{\delta E^j_{\rm nuc}}{\delta X_0}\frac{\delta E^j_{\rm nuc}}{\delta X_1} m_n^{\oplus}n_{n}m_p^{\oplus}n_{p}\mathcal{Y}_{n} \mathcal{Y}_{p} \right]^{-1}\, .
\end{align}
Substituting Eq.~\eqref{eq:Iq-vs-Vq} into \eqref{eq:Veff}, the entrainment matrix can be readily obtained from Eq.~\eqref{eq:Mass-Current-Def2}: 
\beqy\label{eq:entrainment-matrix-Rhonn-final}
\rho_{nn}(T,\pmb{\Veff})=\rho_n \left(1-\mathcal{Y}_{n}\right)\left(\frac{m}{m_n^{\oplus}}+\frac{\mathcal{I}_{nn}}{\hbar}\right)\, ,
\eeqy
\beqy\label{eq:entrainment-matrix-Rhopp-final}
\rho_{pp}(T,\pmb{\Veff})=\rho_p \left(1-\mathcal{Y}_{p}\right)\left(\frac{m}{m_p^{\oplus}}+\frac{\mathcal{I}_{pp}}{\hbar}\right)\, ,
\eeqy
\beqy\label{eq:entrainment-matrix-RhonpRhopn-final}
\rho_{np}(T,\pmb{\Veff})=\rho_n \left(1-\mathcal{Y}_{n}\right)\frac{\mathcal{I}_{np}}{\hbar}\, , \qquad\qquad \rho_{pn}(T,\pmb{\Veff})=\rho_p \left(1-\mathcal{Y}_{p}\right)\frac{\mathcal{I}_{pn}}{\hbar}\, .
\eeqy
Using Eqs.~\eqref{eq:Inp-def} and \eqref{eq:Ipn-def}, one can notice that the entrainment matrix is manifestly symmetric (i.e. $\rho_{np}=\rho_{pn}$). Let us emphasize that Eqs.~\eqref{eq:entrainment-matrix-Rhonn-final}-\eqref{eq:entrainment-matrix-RhonpRhopn-final} give the \emph{exact} expression for the entrainment matrix within the TDHFB theory in homogeneous nuclear matter. No approximation has been made at this point. In particular, the full dependence of the entrainment matrix elements on the currents is taken into account. 

If the currents are small enough such that $\E_{k}>0$, the quasiparticle distributions~\eqref{eq:QuasiparticleDistribution} vanish at $T=0$, hence also the functions $\mathcal{Y}_{q}$, as can be seen from Eq.~\eqref{eq:def-calYq}. In this regime, Eqs.~\eqref{eq:entrainment-matrix-Rhonn-final}-\eqref{eq:entrainment-matrix-RhonpRhopn-final} reduce to the expressions we derived earlier
ignoring nuclear pairing
within the time-dependent Hartree-Fock (TDHF)  equations~\cite{ChamelAllard2019}, namely 
\beqy
\rho^{\text{TDHF}}_{nn}=\rho_n\left[1+ \frac{2}{\hbar^2}\left(\frac{\delta E^j_{\rm nuc}}{\delta X_0}-\frac{\delta E^j_{\rm nuc}}{\delta X_1}\right)\rho_p\right]\, , 
\eeqy 
\beqy
\rho^{\text{TDHF}}_{pp}=\rho_p\left[1+ \frac{2}{\hbar^2}\left(\frac{\delta E^j_{\rm nuc}}{\delta X_0}-\frac{\delta E^j_{\rm nuc}}{\delta X_1}\right)\rho_n\right]\, , 
\eeqy 
\beqy
\rho^{\text{TDHF}}_{np}=\rho^{\text{TDHF}}_{pn}=\rho_n\rho_p\frac{2}{\hbar^2}\left(\frac{\delta E^j_{\rm nuc}}{\delta X_1}-\frac{\delta E^j_{\rm nuc}}{\delta X_0}\right)\, .
\eeqy 
This allows us to rewrite  Eq.~\eqref{eq:Mass-Current-Def2} into the following alternative form
\beqy\label{eq:Mass-Current-Def3}
\pmb{\rho_q}=\sum_{q'}\rho_{qq'}^{\text{TDHF}}\left(\pmb{V_{q'}}-\frac{m_{q'}^{\oplus}}{m}\mathcal{Y}_{q'}\pmb{\mathbb{V}_{q'}}\right)\, .
\eeqy

Finally, let us remark that the relativistic entrainment matrix introduced in Ref.~\cite{gusakov2006} can be directly calculated  from Eqs.~\eqref{eq:entrainment-matrix-Rhonn-final}, \eqref{eq:entrainment-matrix-Rhopp-final}, \eqref{eq:entrainment-matrix-RhonpRhopn-final} using their Eq.~(17): 
\beqy
Y_{nn} &=& \frac{\rho_{nn}-\rho_{np}(\lambda_p/(mc^2))}{(mc)^2 (1+\lambda_n/(mc^2))} \, , \label{eq.Ynn} \\ 
Y_{pp} &=& \frac{\rho_{pp}-\rho_{np}(\lambda_n/(mc^2))}{(mc)^2 (1+\lambda_p/(mc^2))} \, , \label{eq.Ypp} \\ 
Y_{np} &=& Y_{pn}= \frac{\rho_{np}}{(mc)^2}\, . \label{eq.Ynp}
\eeqy

\subsection{Landau's approximations}
\label{sec:Fermi-liquid-approx}

In previous studies of entrainment effects~\cite{gusakov2005,leinson2018}, the mass current was defined as 
\beqy\label{eq:GusakovMassCurrent}
\pmb{\rho_q}=\frac{m}{V}\sum_{k}n_q^{kk}\frac{1}{\hbar}\pmb{\nabla_k}\xi^{(q)}_{k}
\eeqy
($\pmb{\rho_q}$, $n_q^{kk}$ and $\xi^{(q)}_{k}$ were, respectively denoted by $\pmb{J_q}$, $\mathcal{N}^{(q)}_{\pmb{k}+\pmb{Q_q}}$ and $H^{(q)}_{\pmb{k}+\pmb{Q_q}}$, in Ref.~\cite{gusakov2005}, and by $\pmb{j_a}$, $\tilde{n}_{\pmb{k}}^{(a)}$ and $\tilde{\varepsilon}_{\pmb{k}}^{(a)}$, in Ref.~\cite{leinson2018}). Note that $(1/\hbar)\pmb{\nabla_k}\xi^{(q)}_{k}$ represents the group velocity of the single-particle state $k$. 
Using Eqs.~\eqref{eq:xi} and \eqref{eq:Density}, Eq.~\eqref{eq:GusakovMassCurrent} can be expressed as
\begin{align}\label{eq:GusakovJ}
\pmb{\rho_q}&=\frac{m}{m_q^{\oplus}}\hbar\Biggl\{\frac{1}{V}\sum_{k}\left(\pmb{k}+\pmb{Q_q}\right)\left[|\U_{kk}|^2 \f_k+|\V_{k\bar{k}}|^2\left(1-\f_{\bar{k}}\right) \right]\Biggr\}\notag\\
&\qquad\qquad+m\Biggl\{\frac{1}{V}\sum_{k}\left[|\U_{kk}|^2 \f_k+|\V_{k\bar{k}}|^2\left(1-\f_{\bar{k}}\right) \right]\Biggr\}\frac{\pmb{I_q}}{\hbar}\, .
\end{align}
It can be seen from Eqs.~\eqref{eq:Density} and~\eqref{eq:MomentumDensity} that Eq~\eqref{eq:GusakovJ} coincides with our general expression (\ref{eq:mass-current}) of the mass current.  
To compare our results to those obtained earlier within Landau's theory of Fermi liquids, we assume that the superfluid velocities $V_q$ are small compared to the corresponding Fermi velocities $V_{\textrm{F}q}\equiv \hbar k_{\textrm{F}q}/m_q^\oplus$, where $k_{\textrm{F}q}=(3 \pi^2 n_q)^{1/3}$ denotes the Fermi wave number. 
We thus expand the quasiparticle energy~(\ref{eq:QuasiparticleEnergy}) 
as follows: 
\beqy\label{eq:QuasipartEnergy-Lin}
\E_k\approx \Etilde_{\pmb{k}}+\hbar\pmb{k}\cdot\pmb{\Veff}\, ,
\eeqy
\beqy
\Etilde_{\pmb{k}}\equiv\sqrt{\sqepstilde_{\pmb{k}} + \Dtilde^2}= \Etilde_{-\pmb{k}} \, ,
\eeqy
where the single-particle energy $\eps_{\pmb{k}}$ is now evaluated in the static ground state ignoring any dependence on the pairing gaps, and expanding linearly around the Fermi surface
\beqy\label{eq:Epsilonk-def}
\epstilde_{\pmb{k}}\equiv \hbar V_{\text{F}q}(k-k_{\text{F}q})\, ,
\eeqy
which coincides with Eq.~(6) of Ref.~\cite{leinson2018}. In this expression, the reduced chemical potential defined by $\mu_q \equiv \lambda_q - U_q$ has been replaced by the Fermi energy at $T=0$, namely $\mu_q \approx \hbar^2 k_{\text{F}q}^2/(2m_q^\oplus)$ (in general, $\mu_q$ depends on the temperature, the gaps and the currents).  
The quasiparticle distribution function thus becomes
\beqy\label{eq:LinearizedQuasiparticle}
f_k\approx \frac{1}{2}\Biggl\{1-\tanh\left[\frac{\beta}{2}\left( \Etilde_{\pmb{k}} + \hbar \pmb{k}\cdot\pmb{\Veff} \right)\right]\Biggr\}\equiv \fPlus \, .
\eeqy
Similarly, $f_{\bar k} \approx \fMoins$. In the continuum limit, replacing the discrete summation over $\pmb{k}$ by an integral, the function $\mathcal{Y}_q (T,\pmb{\Veff})$ introduced in Eq.~\eqref{eq:def-calYq} thus reads
\begin{equation}
\mathcal{Y}_q (T,\pmb{\Veff})\approx-\frac{\hbar}{\breve{m}_q^{\oplus}n_{q}\Veff}\int \frac{d\Omega_{\pmb{k}}}{4\pi} \int_{-\infty}^{+\infty} \text{d}\epstilde_{\pmb k} \mathcal{D}(\epstilde_{\pmb k}) \left(\fPlus - \fMoins\right) \left( \frac{\epstilde_{\pmb k}}{\hbar V_{\text{F}q}} +k_{\text{F}q}\right)\cos\theta_{\pmb{k}}\, ,
\end{equation} 
where the first integration is over the solid angle in $\pmb{k}$-space, $\theta_{\pmb{k}}$ is the angle between $\pmb{k}$ and $\pmb{\Veff}$, and $\mathcal{D}(\epstilde_{\pmb k})$ is the level density\footnote{Note that $\mathcal{D}(\epstilde_{\pmb k})$ denotes the level density per one spin state since summation over spins is already taken into account in Eq.~\eqref{eq:def-calYq}}. In the Landau's theory, the level density is further assumed to be constant and approximated by its value on the Fermi surface, namely $\mathcal{D}(\epstilde_{\pmb k})\approx  \mathcal{D}(0)= k_{\text{F}q}\breve{m}_q^{\oplus}/(2\pi^2\hbar^2)$. Moreover, the term $\epstilde_{\pmb k}/(\hbar V_{\text{F}q})$ is also evaluated on the Fermi surface, and therefore is dropped. With all these approximations, the function $\mathcal{Y}_q$ finally reduces to Eq.~(70) of Ref.~\cite{leinson2018} (where $\mathcal{Y}$ was denoted by $\tilde{\Phi}$): 
\beqy\label{eq:Yq-Landau}
\mathcal{Y}_q(T,\pmb{\Veff}) \approx -\frac{3}{\hbar k_{F_q}\Veff}\int \frac{{\rm d}\Omega_{\pmb{k}}}{4\pi}\int_0^{+\infty} d\epstilde_{\pmb k} \left(\fPlus - \fMoins\right) \cos\theta_{\pmb{k}}\, .
\eeqy 
Note that a factor of 2 was absorbed by integrating over half the energy domain since the function to integrate is invariant under the change $\epstilde_{\pmb k}\rightarrow -\epstilde_{\pmb k}$. 
Similar approximations for the gap equations~\eqref{eq:GapEquation} yield 
\beqy\label{eq:GapEquation2}
\breve{\Delta}_q \approx 4\frac{\delta E}{\delta\vert \widetilde{n}_q\vert^2}\mathcal{D}(0) \int \frac{d\Omega_{\pmb{k}}}{4\pi}\int_0^{+\infty} d\epstilde_{\pmb k} \frac{\breve{\Delta}_q}{\Etilde_{\pmb{k}}}(\fPlus + \fMoins-1) \, .
\eeqy
Let us remark that the gap equations are thus decoupled from the particle number conservation~\eqref{eq:Density} since the reduced chemical potentials are approximated by the corresponding Fermi energies at $T=0$. Further assuming $\breve{\Delta}_q\ll \mu_q$, 
Eq.~\eqref{eq:GapEquation2} can be expressed as~\cite{leinson2018}
\beqy
\ln\left(\frac{\breve{\Delta}^{(0)}_q}{\breve{\Delta}_q}\right)=\int\frac{{\rm d}\Omega_{\pmb{k}}}{4\pi}\int_0^{+\infty} {\rm d}\epstilde_{\pmb k} \frac{\fPlus + \fMoins}{\Etilde_{\pmb{k}} }\, ,
\eeqy
where $\breve{\Delta}^{(0)}_q$ denotes the solution of Eq.~\eqref{eq:GapEquation2} at $T=0$ in the absence of currents. 

Substituting Eq.~\eqref{eq:Yq-Landau} in Eqs.~\eqref{eq:Inn-def}-\eqref{eq:Theta-def}, and evaluating all other quantities for the superfluids at rest,  the entrainment matrix~\eqref{eq:entrainment-matrix-Rhonn-final}-\eqref{eq:entrainment-matrix-RhonpRhopn-final} can be recast in a form similar to Eq.~(30) of Ref.~\cite{leinson2018}: 
\beqy\label{eq:entrainment-matrix-Landau}
\rho_{qq'}(T,\pmb{\Veff})=\rho_q \left(1-\mathcal{Y}_{q}\right)\gamma_{qq'} =\rho_{q'} \left(1-\mathcal{Y}_{q'}\right)\gamma_{q'q}\, , 
\eeqy
\beqy\label{eq:GammaNN-def}
\gamma_{nn}=\frac{m}{\breve{m}_{n}^{\oplus}S}\left[\left(1+\frac{\mathcal{F}_1^{nn}}{3}\right)\left(1+\frac{\mathcal{F}_1^{pp}}{3}\mathcal{Y}_{p}\right)-\left(\frac{\mathcal{F}_1^{np}}{3}\right)^2\mathcal{Y}_{p}\right] \, , 
\eeqy
\beqy\label{eq:GammaPP-def}
\gamma_{pp}=\frac{m}{\breve{m}_{p}^{\oplus}S}\left[\left(1+\frac{\mathcal{F}_1^{pp}}{3}\right)\left(1+\frac{\mathcal{F}_1^{nn}}{3}\mathcal{Y}_{n}\right)-\left(\frac{\mathcal{F}_1^{pn}}{3}\right)^2\mathcal{Y}_{n}\right] \, , 
\eeqy
\beqy\label{eq:GammaNP-def}
\gamma_{np}=\frac{m}{3S\sqrt{\breve{m}_n^{\oplus}\breve{m}_{p}^{\oplus}}}\left(\frac{n_{p}}{n_{n}}\right)^{1/2}\mathcal{F}_1^{np}\left(1-\mathcal{Y}_{p}\right)\, ,
\eeqy
\beqy\label{eq:GammaPN-def}
\gamma_{pn}=\frac{m}{3S\sqrt{\breve{m}_p^{\oplus}\breve{m}_{n}^{\oplus}}}\left(\frac{n_{n}}{n_{p}}\right)^{1/2}\mathcal{F}_1^{pn}\left(1-\mathcal{Y}_{n}\right)\, ,
\eeqy
where $\mathcal{F}_1^{qq'}$ denotes the dimensionless $\ell=1$ Landau parameter 
(derivatives are calculated in the absence of currents)
\beqy
\mathcal{F}_1^{qq'}=\frac{6}{\hbar^2}\left[\left(1-2\delta_{qq'}\right)\frac{\delta E^j_{\textrm{nuc}}}{\delta X_1}\biggr\vert_0-\frac{\delta E^j_{\textrm{nuc}}}{\delta X_0}\biggr\vert_0\right]\sqrt{\breve{m}_q^{\oplus}n_{q} \breve{m}_{q'}^{\oplus}n_{q'}}\, ,
\eeqy
and the function $S$ is given by   
\begin{align}\label{eq:S-def}
S &=\Theta^{-1}= \left(1+\frac{\mathcal{F}^{nn}_1}{3}\mathcal{Y}_n \right)\left(1+\frac{\mathcal{F}^{pp}_1}{3}\mathcal{Y}_{p}\right)-\left(\frac{\mathcal{F}^{np}_1}{3}\right)^2\mathcal{Y}_n\mathcal{Y}_{p}\, . 
\end{align}

Simplifying the exact solution~\eqref{eq:entrainment-matrix-Rhonn-final}-\eqref{eq:entrainment-matrix-RhonpRhopn-final} using Landau's approximations, we have thus recovered the entrainment matrix 
derived earlier in Refs.~\cite{leinson2018,gusakov2005}. It should be stressed that the above formulas do not account for the full dependence of the entrainment matrix on the currents unlike Eqs.~\eqref{eq:entrainment-matrix-Rhonn-final}-\eqref{eq:entrainment-matrix-RhonpRhopn-final}. In particular, the nonlinear effects contained in the single-particles energies have been ignored, compare Eqs.~\eqref{eq:Epsilon} and \eqref{eq:Epsilonk-def} recalling that the mean fields themselves have a highly nonlinear dependence on the currents.

\section{Conclusions}

We have studied the dynamics of nuclear superfluid systems at finite temperatures and finite currents in the framework of the self-consistent time-dependent nuclear-energy density functional theory. Considering the TDHFB equations in coordinate space, we have derived general expressions for the local nucleon mass currents~\eqref{eq:mass-current} and local superfluid velocities~\eqref{eq:super-vel-def}, which are valid for both homogeneous systems (such as the outer core of neutron stars) and inhomogeneous systems (such as the crust of neutron stars, atomic nuclei and vortices). Remarkably, the mass currents are found to have the same formal expressions at any temperature, and coincide with the ones we derived earlier in the absence of pairing from the TDHF equations at zero temperature~\cite{ChamelAllard2019}. 

Focusing on homogeneous neutron-proton superfluid mixtures, we have shown that the TDHFB equations can be solved exactly for arbitrary temperatures and currents. Using this solution, we have been able to express   the Andreev-Bashkin mutual entrainment matrix in analytic form. 
The formal simplicity of our expression~\eqref{eq:entrainment-matrix-Rhonn-final}-\eqref{eq:entrainment-matrix-RhonpRhopn-final} are however deceptive: the entrainment coupling coefficients depend themselves on the currents in a very complicated way through the self-consistency of the TDHFB equations. We have explicitly demonstrated that our 
expression reduces to that obtained earlier in Ref.~\cite{leinson2018} within  Landau's theory. 

Our formulas are applicable to a large class of nuclear energy-density functionals. These include the Brussels-Montreal functionals based on generalized Skyrme effective interactions, for which unified equations of state for all regions of neutron stars have been calculated~\cite{potekhin2013,mutafchieva2019, pearson2018,pearson2020}, thus paving the way for a fully consistent microscopic treatment of the dynamics of superfluid neutron stars. The formalism we have developed here in the nuclear context may also be easily transposed to the less exotic kinds of superfluids studied in terrestrial laboratories and described by similar TDHFB equations.

The relativistic entrainment matrix introduced in Ref.~\cite{gusakov2006}  is given by Eqs.~\eqref{eq.Ynn}, \eqref{eq.Ypp}, and \eqref{eq.Ynp}. However, it would be worth carrying out the same analysis within a fully relativistic self-consistent microscopic framework, using the relativistic finite temperature HFB method~\cite{jiajie2015}.

\begin{acknowledgments}
The authors thank T. Duguet and J. Sauls for valuable discussions. This work was financially  supported by the Fonds de la Recherche Scientifique (Belgium) under Grant No. PDR T.004320. This work was also partially supported by the European Cooperation in Science and Technology action (EU) CA16214. 
\end{acknowledgments}

\appendix
\section{Continuity equations}
\label{app:continuity}

Making use of the completeness relations 
\beqy
\sum_i \varphi^{(q)}_i(\pmb{r},\sigma)^*\varphi^{(q)}_i(\pmb{r^\prime},\sigma^\prime)=\delta(\pmb{r}-\pmb{r^\prime})\delta_{\sigma\sigma^\prime}\, , 
\eeqy
and using Eqs.~\eqref{eq:Hamiltonian-matrix2} and \eqref{eq:pairing-matrix2}, we can demonstrate the following identities: 
\beqy 
\label{eq:Hamiltonian-matrix3}
\sum_{i,j} \h^{ij}(t)\varphi^{(q)}_i(\pmb{r},\sigma)\varphi^{(q)}_j(\pmb{r^\prime},\sigma^\prime)^*=\h(\pmb{r},t)\delta(\pmb{r}-\pmb{r^\prime})\delta_{\sigma\sigma^\prime}\, ,
\eeqy 
\beqy\label{eq:pairing-matrix3}
\sum_{i,j} \D^{ij}(t)\varphi^{(q)}_i(\pmb{r},\sigma)(-\sigma^\prime)\varphi^{(q)}_j(\pmb{r^\prime},-\sigma^\prime)=\D(\pmb{r},t)\delta(\pmb{r}-\pmb{r^\prime})\delta_{\sigma\sigma^\prime}\, .
\eeqy 

Multiplying Eq.~\eqref{eq:TDHFB1} by $\varphi^{(q)}_i(\pmb{r},\sigma)\varphi^{(q)}_j(\pmb{r^\prime},\sigma^\prime)^*$ and summing over indices $i$ and $j$ yields 
\begin{align}
\label{eq:continuity-term1}
&i\hbar \frac{\partial}{\partial t} \sum_{i,j}  n_q^{ij} \varphi^{(q)}_i(\pmb{r},\sigma)\varphi^{(q)}_j(\pmb{r^\prime},\sigma^\prime)^*=\\
\label{eq:continuity-term2}
&\qquad\qquad\sum_{i,j,k}
 \left(\h^{ik}\varphi^{(q)}_i(\pmb{r},\sigma)\varphi^{(q)}_j(\pmb{r^\prime},\sigma^\prime)^*n_q^{kj}-n_q^{ik}\varphi^{(q)}_i(\pmb{r},\sigma)\varphi^{(q)}_j(\pmb{r^\prime},\sigma^\prime)^*\h^{kj}\right)\\ 
\label{eq:continuity-term3}
&\qquad\quad +\sum_{i,j,k}\left(\kappa_q^{ik}\varphi^{(q)}_i(\pmb{r},\sigma)\varphi^{(q)}_j(\pmb{r^\prime},\sigma^\prime)^*\D^{kj*} -\D^{ik}\varphi^{(q)}_i(\pmb{r},\sigma)\varphi^{(q)}_j(\pmb{r^\prime},\sigma^\prime)^*\kappa_q^{kj*}\right) \, .
\end{align}
The summation in \eqref{eq:continuity-term1} yields the density matrix $n_q(\pmb{r}, \sigma; \pmb{r^\prime}, \sigma^\prime; t)$, as can be seen from Eq.~\eqref{eq:DensityMatrixCoordinateSpaceDef}. The next two summations in Eq.~\eqref{eq:continuity-term2} can be simplified using Eq.~\eqref{eq:Hamiltonian-matrix3} and the orthonormality property of the single-particle wavefunctions
\beqy
\sum_\sigma \int {\rm d}^3\pmb{r}\, \varphi^{(q)}_i(\pmb{r},\sigma)\varphi^{(q)}_j(\pmb{r},\sigma)^* = \delta_{ij}\, .
\eeqy 
We can thus write 
\begin{align}
\sum_{i,j,k}&\h^{ik}\varphi^{(q)}_i(\pmb{r},\sigma)\varphi^{(q)}_j(\pmb{r^\prime},\sigma^\prime)^* n_q^{kj}=\sum_{i,j,k,l}\h^{ik}\varphi^{(q)}_i(\pmb{r},\sigma)\delta_{kl}\varphi^{(q)}_j(\pmb{r^\prime},\sigma^\prime)^* \, n_q^{lj}\nonumber \\
&=\sum_{\sigma^{\prime\prime}}\int\,{\rm d}^3\pmb{r^{\prime\prime}}\, 
\sum_{i,j,k,l}
 \h^{ik}\varphi^{(q)}_i(\pmb{r},\sigma)\varphi^{(q)}_k(\pmb{r^{\prime\prime}},\sigma^{\prime\prime})^*\varphi^{(q)}_l(\pmb{r^{\prime\prime}},\sigma^{\prime\prime})\varphi^{(q)}_j(\pmb{r^\prime},\sigma^\prime)^* n_q^{lj} \nonumber \\
&=\h(\pmb{r},t)\sum_{\sigma^{\prime\prime}}\int\,{\rm d}^3\pmb{r^{\prime\prime}}\,  \delta(\pmb{r}-\pmb{r^{\prime\prime}})\delta_{\sigma\sigma^{\prime\prime}}\, n_q(\pmb{r^{\prime\prime}}, \sigma^{\prime\prime}; \pmb{r^\prime}, \sigma^\prime; t)\nonumber \\ 
&=\h(\pmb{r},t)\, n_q(\pmb{r}, \sigma; \pmb{r^\prime}, \sigma^\prime; t)\, .
\end{align}
Similarly, we have
\begin{align}
\sum_{i,j,k}&n_q^{ik}\varphi^{(q)}_i(\pmb{r},\sigma)\varphi^{(q)}_j(\pmb{r^\prime},\sigma^\prime)^*\h^{kj}=\sum_{i,j,k,l} n_q^{ik}\varphi^{(q)}_i(\pmb{r},\sigma)\delta_{kl}\varphi^{(q)}_j(\pmb{r^\prime},\sigma^\prime)^*\h^{lj}\nonumber \\
&=\sum_{\sigma^{\prime\prime}}\int\,{\rm d}^3\pmb{r^{\prime\prime}}\, 
\sum_{i,j,k,l} n_q^{ik}\varphi^{(q)}_i(\pmb{r},\sigma)\varphi^{(q)}_k(\pmb{r^{\prime\prime}},\sigma^{\prime\prime})^* \varphi^{(q)}_l(\pmb{r^{\prime\prime}},\sigma^{\prime\prime})\varphi^{(q)}_j(\pmb{r^\prime},\sigma^\prime)^*\h^{lj}\nonumber \\
& =\sum_{\sigma^{\prime\prime}}\int\,{\rm d}^3\pmb{r^{\prime\prime}}\, n_q(\pmb{r}, \sigma; \pmb{r^{\prime\prime}}, \sigma^{\prime\prime}; t) \h(\pmb{r^{\prime\prime}},t)\delta(\pmb{r^{\prime\prime}}-\pmb{r^\prime})\delta_{\sigma^\prime\sigma^{\prime\prime}} \nonumber \\ 
\label{eq:continuity-term2b}
& =\int\,{\rm d}^3\pmb{r^{\prime\prime}}\, n_q(\pmb{r}, \sigma; \pmb{r^{\prime\prime}}, \sigma^{\prime}; t) \h(\pmb{r^{\prime\prime}},t)\delta(\pmb{r^{\prime\prime}}-\pmb{r^\prime})\, .
\end{align}
Let us recall that $\h(\pmb{r^{\prime\prime}},t)$ involves the operator $\pmb{\nabla^{\prime\prime}}$ so that it cannot be factored out of the integral. However, the hermiticity of the Hamiltonian matrix, $\h^{ij}=\h^{ji*}$, implies the following identity~\cite{ChamelAllard2019}: 
\beqy
\h(\pmb{r^{\prime\prime}},t)\delta(\pmb{r^{\prime\prime}}-\pmb{r^\prime})=\h(\pmb{r^{\prime}},t)^*\delta(\pmb{r^{\prime\prime}}-\pmb{r^\prime})\, .
\eeqy 
Finally, Eq.~\eqref{eq:continuity-term2b} becomes 
\begin{align}
&\sum_{i,j,k}n_q^{ik}\varphi^{(q)}_i(\pmb{r},\sigma)\varphi^{(q)}_j(\pmb{r^\prime},\sigma^\prime)^*\h^{kj}=\h(\pmb{r^{\prime}},t)^* n_q(\pmb{r}, \sigma; \pmb{r^{\prime}}, \sigma^{\prime}; t) \, .
\end{align}
Likewise, the two summations in Eq.~\eqref{eq:continuity-term3} can be expressed as follows using Eqs.~\eqref{eq:AbnormalDensityMatrixCoordinateSpaceDef} and \eqref{eq:pairing-matrix3}: 
\begin{align}
\sum_{i,j,k}&\kappa_q^{ik}\varphi^{(q)}_i(\pmb{r},\sigma)\varphi^{(q)}_j(\pmb{r^\prime},\sigma^\prime)^*\D^{kj*}=\sum_{i,j,k,l}\kappa_q^{ik}\varphi^{(q)}_i(\pmb{r},\sigma)\delta_{kl}\varphi^{(q)}_j(\pmb{r^\prime},\sigma^\prime)^*\D^{lj*} \nonumber \\
&=\sum_{\sigma^{\prime\prime}}\int\,{\rm d}^3\pmb{r^{\prime\prime}}\, \sum_{i,j,k,l}\kappa_q^{ik}\varphi^{(q)}_i(\pmb{r},\sigma)\varphi^{(q)}_k(\pmb{r^{\prime\prime}},-\sigma^{\prime\prime})(-\sigma^{\prime\prime})(-\sigma^{\prime\prime})\varphi^{(q)}_l(\pmb{r^{\prime\prime}},-\sigma^{\prime\prime})^* \varphi^{(q)}_j(\pmb{r^\prime},\sigma^\prime)^*\D^{lj*} \nonumber \\ 
&=-\D(\pmb{r^\prime},t)^*\sum_{\sigma^{\prime\prime}}\int\,{\rm d}^3\pmb{r^{\prime\prime}}\, \widetilde{n}_q(\pmb{r}, \sigma; \pmb{r^{\prime\prime}}, \sigma^{\prime\prime}; t)\delta(\pmb{r^\prime}-\pmb{r^{\prime\prime}})\delta_{\sigma^\prime\sigma^{\prime\prime}} \nonumber \\
&=-\widetilde{n}_q(\pmb{r}, \sigma; \pmb{r^{\prime}}, \sigma^{\prime}; t)\D(\pmb{r^\prime},t)^*\, , 
\end{align}
\begin{align}
\sum_{i,j,k}&\D^{ik}\varphi^{(q)}_i(\pmb{r},\sigma)\varphi^{(q)}_j(\pmb{r^\prime},\sigma^\prime)^*\kappa_q^{kj*}=\sum_{i,j,k,l}\D^{ik}\varphi^{(q)}_i(\pmb{r},\sigma)\delta_{kl}\varphi^{(q)}_j(\pmb{r^\prime},\sigma^\prime)^*\kappa_q^{lj*} \nonumber \\ 
&=\sum_{\sigma^{\prime\prime}}\int\,{\rm d}^3\pmb{r^{\prime\prime}}\,\sum_{i,j,k,l}\D^{ik}\varphi^{(q)}_i(\pmb{r},\sigma)\varphi^{(q)}_k(\pmb{r^{\prime\prime}},-\sigma^{\prime\prime})(-\sigma^{\prime\prime})(-\sigma^{\prime\prime})\varphi^{(q)}_l(\pmb{r^{\prime\prime}},-\sigma^{\prime\prime})^*\varphi^{(q)}_j(\pmb{r^\prime},\sigma^\prime)^*\kappa_q^{lj*}  \nonumber \\ 
&=-\D(\pmb{r},t)\sum_{\sigma^{\prime\prime}}\int\,{\rm d}^3\pmb{r^{\prime\prime}}\,\delta(\pmb{r}-\pmb{r^{\prime\prime}})\delta_{\sigma\sigma^{\prime\prime}}\widetilde{n}_q(\pmb{r^\prime}, \sigma^\prime; \pmb{r^{\prime\prime}}, \sigma^{\prime\prime}; t)^*\nonumber \\ 
&=-\D(\pmb{r},t)\widetilde{n}_q(\pmb{r^\prime}, \sigma^\prime; \pmb{r}, \sigma; t)^*\, .
\end{align}
Collecting terms, multiplying by $\delta(\pmb{r}-\pmb{r^\prime})$, integrating over $\pmb{r^\prime}$ and summing over spins $\sigma^\prime=\sigma$ lead to 
\begin{align}
i\hbar \frac{\partial n_q(\pmb{r};t)}{\partial t}
=&\sum_{\sigma}\int\,{\rm d}^3\pmb{r^{\prime}}\,\delta(\pmb{r}-\pmb{r'}) \left[\h(\pmb{r},t) n_q(\pmb{r}, \sigma; \pmb{r^{\prime}}, \sigma; t)- \h(\pmb{r^{\prime}},t)^* n_q(\pmb{r}, \sigma; \pmb{r^{\prime}}, \sigma; t)\right] \nonumber \\
&+\D(\pmb{r},t)\widetilde{n}_q(\pmb{r}; t)^* - \widetilde{n}_q(\pmb{r}; t)\D(\pmb{r},t)^* \, .
\end{align}
As shown in Ref.~\cite{ChamelAllard2019}, the integral can be equivalently expressed as the divergence of a particle flux. The last two terms cancel each other if the pairing potential is calculated self-consistently from Eq.~\eqref{eq:pair-pot2}. Finally, multiplying by $m/(i\hbar)$ leads to the same  Eq.~\eqref{eq:continuity-final} as previously derived in Ref.~\cite{ChamelAllard2019} ignoring pairing.

\bibliography{references.bib}

\end{document}